\documentclass[useAMS,usenatbib]{mn2e}
\usepackage{multirow}
\usepackage{epsfig}
\usepackage{subfigure}

\makeatother

\title[New Compton-thick AGN]{New Compton-thick AGN in the circumnuclear H$_2$O maser hosts UGC~3789 and NGC~6264}
\author[]{P. Castangia$^1$\thanks{E-mail: pcastang@oa-cagliari.inaf.it}, F. Panessa$^{2, 3}$, C. Henkel
$^{4, 5}$, M. Kadler$^6$, A. Tarchi$^1$ \\
$^1$INAF-Osservatorio Astronomico di Cagliari, Via della Scienza, 09047 Selargius (CA), Italy\\ 
$^2$Istituto di Astrofisica e Planetologia Spaziali di Roma (IAPS-INAF), 
via del Fosso del Cavaliere 100, 00133 Roma, Italy\\
$^3$Dipartimento di Matematica e Fisica, Universit\`a degli Studi Roma Tre, Via della Vasca Navale 84, I-00146 Roma, Italy\\
$^4$Max-Planck-Institut f\"{u}r Radioastronomie, Auf dem H\"{u}gel 69, 53121 Bonn, Germany\\
$^5$Astronomy Dept., Kind Abdulaziz University, P.O. Box 80203, Jeddah, Saudi Arabia\\
$^6$Lehrstuhl f\"ur Astronomie, Universit\"at W\"urzburg, Emil-Fischer-Straße 31, 97074, W\"urzburg, Germany\\
}
\begin{document}

\date{}

\pagerange{\pageref{firstpage}--\pageref{lastpage}} \pubyear{2002}

\maketitle

\label{firstpage}

\begin{abstract}
Large column densities, derived from X-ray studies, are typically measured towards AGN hosting water masers, especially when the H$_2$O emission is associated with the nuclear accretion disk. In addition, possible correlations between the intrinsic X-ray luminosity and the characteristics of the H$_2$O maser emission have been put forward that, however, require confirmation. We have performed high-sensitivity {\it XMM-Newton} observations of a sample of five H$_2$O maser sources confidently detected in our ongoing survey with the {\it Swift} satellite of all known water masers in AGN, in order to obtain detailed X-ray information of these promising targets and to set up a systematic detailed study of the X-ray/H$_2$O-maser relation in AGN. 
For three galaxies, NGC\,613, VII\,Zw\,73, and IRAS\,16288+3929, the amount of intrinsic absorption has been estimated, indicating column densities of 4--6$\times$10$^{23}$\,cm$^{-2}$. For UGC\,3789 and NGC\,6264 (the two confirmed disk-maser galaxies in our sample), column densities in excess of 1$\times$10$^{24}$\,cm$^{-2}$ are inferred from the large $EW$ of the Fe K$\alpha$ line. By adding our results to those obtained in past similar studies, we find that the percentage of water masers sources that host highly-obscured ($N_{\rm H}>10^{23}$\,cm$^{-2}$) and Compton-thick AGN is 96\% (45/47) and 57\% (27/47), respectively. In addition, 86\%, 18/21 of disk maser galaxies host Compton-thick AGN. The correlation between the galaxies' bolometric luminosity and accretion disk radius, suggested in previous works, is also confirmed. 
\end{abstract}


\section{Introduction}\label{sect:intro}

The number density and physical properties (e.\,g. mass and spin) of active galactic nuclei (AGN) in the local Universe is fundamental to understand the birth and growth of supermassive black holes (SMBH) \citep[e.\,g.][]{marconi04}. 
However, there is evidence that a significant fraction of AGN in the local Universe may be missed even in the deepest optical surveys \citep{goulding2010} which are biased against gas-rich and dust-obscured objects. In fact, as the amounts of gas and dust along the line of sight increases, the search and study of AGN in the optical, UV, and soft (below 2\,keV) X-ray bands becomes more and more difficult because the radiation emitted by the central engine at these frequencies is heavily absorbed by the intervening medium. The radio emission from the 22\,GHz H$_2$O maser line can pass through the absorbing matter and constitutes one of the most powerful tools to study the structure and dynamics of the molecular gas in the inner few parsec of obscured AGN (for recent reviews see \citealt{lo05,henkel05,greenhill07,tarchi2012}). Indeed, luminous extragalactic water masers (the so-called ``megamasers'') originate within a few parsec from the nuclear engines, tracing circumnuclear accretion disks \citep[``disk-masers'', e.\,g. UGC~3789:][]{reid09,braatz2010,reid2013}, the inner part of relativistic jets \citep[``jet-masers'', e.\,g. Mrk348:][]{peck03} or nuclear outflows \citep[Circinus:][]{greenhill03}. Very Long Baseline Interferometry (VLBI) and single-dish monitoring studies of water maser sources allow us to determine the geometry of sub-parsec scale accretion disks and the enclosed dynamical masses \citep[e.\,g.][]{greene2010,kuo2011}. Water megamaser observations provide the most precise method to derive black hole masses, especially in the case of obscured AGN, where the optical tracers of the gas dynamics (e.\,g. the broad line clouds) are hidden from direct view. Another fundamental instrument for a detailed investigation of absorbed AGN are X-ray observations at energies $>$2\,keV. In fact, X-ray emission is produced within a few gravitational radii from the black hole \citep[e.\,g.][]{risaliti09}. X-ray spectral analysis yields a measure of the absorbing column density and, hence, of the intrinsic luminosity of the AGN. 
Therefore, the combination of X-ray and H$_2$O maser studies may potentially provide an invaluable means to probe the physics of obscured AGN. 

Initial investigations of H$_2$O megamaser galaxies at X-ray energies revealed that most of them harbour highly absorbed AGN, with column densities, $N_{\rm H}$, larger than $10^{23}$\,cm$^{-2}$ \citep{zhang06}. Studying a sample of 42 masers in AGN, \citet*{greenhill08} reports that 95\% of the objects show $N_{\rm H}>10^{23}$\,cm$^{-2}$ and, most notably, 60\% are Compton-thick ($N_{\rm H}>10^{24}$\,cm$^{-2}$). The fraction of Compton-thick AGN increases (up to 76\%) when the sub-sample of disk-masers (21 objects) is considered. If these results are confirmed, radio surveys to find water maser sources might become an efficient way to identify potential highly obscured AGN \citep{greenhill08}. This would be particularly important for the most absorbed sources. Indeed, the number of Compton-thick AGN discovered so far is surprisingly low compared to the predictions of the X-ray background synthesis models. According to \citet*{gilli07}, for example, the fraction of Compton-thick AGN required to reproduce the 30 keV peak of the cosmic X-ray background is 20--30\%, while even in the deepest hard X-ray surveys (the least biased against nuclear obscuration) the observed fraction of Compton thick AGN is $<10$\% \citep{malizia2012,burlon2011}. Indeed, hard X-ray instruments such as {\it BAT} and {\it INTEGRAL} still miss Compton-thick objects which are intrinsically weak and/or at large distances. Correcting for this bias, the reported intrinsic fraction of Compton-thick AGN becomes $\sim$20\% \citep{malizia09,malizia2012,burlon2011}. Possible correlations between the intrinsic X-ray luminosity and the properties of H$_2$O maser emission (i.\,e. the isotropic line luminosity and the radius of the masing disk) have also been suggested \citep*{kondratko06a,tilak08}. 

In the Compton-thin regime ($N_{\rm H}<10^{24}$\,cm$^{-2}$), the primary continuum emission of the AGN is dimmed but can still be observed in the 2--10\,keV energy range. However, when the absorbing column density exceeds $10^{24}$\,cm$^{-2}$, the direct component is strongly suppressed also at these energies becoming visible only above 10\,keV. The 2--10\,keV X-ray spectra of Compton-thick AGN are dominated by a reflected component that is believed to be the nuclear continuum emission reprocessed by cold material in the putative torus that surrounds the X-ray source in the Unification Model \citep{antonucci93,urry95}. The cold reflecting medium is also predicted to produce a fluorescence iron line at $\sim$6.4\,keV \citep[e.\,g.][and references therein]{matt07}. Being the reflected component much weaker than the direct one, X-ray spectra of Compton-thick sources below 10\,keV do not allow a proper characterization of the nuclear continuum and a direct measure of the absorbing column density to be made. 
Hard X-ray data (in the 10--100\,keV band) allow us to estimate the amount of intrinsic absorption and to distinguish between an heavily absorbed Compton-thin or a Compton-thick object as far as $N_{\rm H}<10^{25}$\,cm$^{-2}$ (``mildly'' Compton-thick sources) \citep[e.\,g.][]{severgnini2011,derosa2012,malizia2012}. However, for larger values of the absorbing column density (``heavily'' Compton-thick objects), X-ray photons are almost completely blocked also at these high energies. For highly absorbed AGN without high-energy data, or when $N_{\rm H}>10^{25}$\,cm$^{-2}$, the presence of Compton-thick matter along the line of sight is inferred through indirect arguments: the equivalent width ($EW$) of the Fe K$\alpha$ line and the comparison of the observed X-ray flux with a more isotropic indicator of the intrinsic brightness of the nucleus, e.\,g. the [O{\sc iii}] optical-line flux arising from the narrow line region \citep*[e.\,g.][]{guainazzi05,panessa06,zhang2010}.

In order to obtain detailed X-ray information on the largest possible number of H$_2$O maser sources, we have performed a survey of all known water masers in AGN using the {\it Swift} satellite \citep{castangia2010}. The data reduction is still on-going, nonetheless, to date, we have detected about 60\% of the targets with the X-Ray Telescope (XRT, 0.2--10\,keV) and increased the number of H$_2$O maser sources with X-ray data by 20\% \citep{castangia2010}. Here we present {\it XMM-Newton} follow-ups of a first sample of five H$_2$O maser sources selected from those galaxies detected by us with {\it Swift/XRT} with a significance $> 5 \sigma$ but whose spectra do not permit an estimate of $N_{\rm H}$, due to the poor counting statistics. The brighest sources have also been detected by the Burst Alert Telescope (BAT) at energies $>$15\,keV. The targets are listed in Table~\ref{table:sou_info}. Two of them, UGC~3789 and NGC~6264, host VLBI confirmed disk-maser emission \citep{reid09,kuo2011,reid2013,kuo2013}. For the other three sources, the origin of the H$_2$O maser emission has not been established yet, although an association with a jet or outflow seems to be more likely \citep*{kondratko06b,castangia08,greenhill09}.  

\section{The {\it XMM-Newton} data reduction} {\it XMM-Newton} observation dates, EPIC CCDs (pn and MOS) effective exposure times and filters used during the observations are shown in Table~\ref{table:obs_info}. The raw observation data files (ODFs) were reduced and analyzed using the standard Science Analysis System (SAS) software package (version 10.0.0) with associated latest calibration files. Data were cleaned through an inspection of the high energy light curves and time intervals with very high background rates were removed. The source spectra and light curves were extracted from a 25$^{\prime\prime}$ circular region centered on the source in order to reduce contamination from surrounding diffuse emission, while background circular regions were chosen to be free from contaminating sources. The {\it arfgen} and {\it rmfgen} SAS tasks were used to create ancillary and response files. Spectral channels were binned to a minimum of 20 counts per bin and spectra were fitted using data from the three CCD detectors (MOS1/2 and pn) simultaneously. We have fixed the pn normalization to 1 and left the MOS1/2 normalizations free to vary in the range 0.95--1.05 to account for the remaining calibration uncertainties in their cross-normalizations. 

\begin{table*}
\caption{\bf Galaxy properties}
\small{
\begin{center}
\begin{tabular}{lcccccccccccc}
\hline
\hline
\multicolumn{1}{c}{Name} &
\multicolumn{1}{c}{Distance} &
\multicolumn{1}{c}{Optical Class} &
\multicolumn{1}{c}{$F_{12}$} &
\multicolumn{1}{c}{$F_{25}$} &
\multicolumn{1}{c}{$F_{60}$} &
\multicolumn{1}{c}{$F_{100}$} &
\multicolumn{1}{c}{$F_{[\rm OIII]}$} &
\multicolumn{1}{c}{$F_{\rm [OIII], corr}$} &
\multicolumn{1}{c}{H$_{\alpha}$/H$_{\beta}$} &
\multicolumn{1}{c}{Ref.} \\
\multicolumn{1}{c}{(1)} &
\multicolumn{1}{c}{(2)} &
\multicolumn{1}{c}{(3)} &
\multicolumn{1}{c}{(4)} &
\multicolumn{1}{c}{(5)} &
\multicolumn{1}{c}{(6)} &
\multicolumn{1}{c}{(7)} &
\multicolumn{1}{c}{(8)} &
\multicolumn{1}{c}{(9)} &
\multicolumn{1}{c}{(10)} &
\multicolumn{1}{c}{(11)} \\
\hline
\hline
NGC~613  & 20  & H{\sc ii}/Sy & 0.74 & 2.08 & 19.6 & 49.1 & 0.17 & 1.91  & 6.80 & Gou09 \\
VIIZw73  & 170 & Sy2          & 0.10 & 0.50 & 1.76 & 2.46 & 0.74 & 0.74  & 2.70 & deG92 \\
UGC~3789 & 46  & Sy2          & 0.11 & 0.34 & 1.65 & 3.47 & 7.00 & 50.62 & 5.88 & Gre10 \\
IRAS~16288+3929 & 125 & Sy2   & $<$0.09 & 0.20 & 0.68 & 0.54 & 0.69 & 1.66 & 4.04 & deG92 \\
NGC~6264 & 140  & Sy2          & --   & --   & --   & --   & 0.29 & 1.63  & 5.39 & Gal96 \\
\hline
\end{tabular}
\end{center}
Notes: (1): Galaxy name. 
(2): Distance in Mpc. 
(3): Optical class. 
(4),(5),(6),(7): Infrared fluxes at 12, 25, 60, and 100$\mu$m in Jy, 
taken from IRAS Point Source and IRAS Faint Source Catalogs. 
(8): Observed [O{\sc iii}]$\lambda$5007 flux, in units of 10$^{-13}$ erg cm$^{-2}$. 
(9): [O{\sc iii}]$\lambda$5007 flux, in units of 10$^{-13}$ erg cm$^{-2}$, corrected for Galactic and intrinsic extinction, following the method in \citet{bassani99}. 
(10): Intensity ratio H$_{\alpha}$/H$_{\beta}$. 
(11): References for the $F_{[\rm OIII]}$ flux and H$_{\alpha}$/H$_{\beta}$. deG92: \citet{degrijp92}. Gal96: \citet{gallego96}. Gou09: \citet{goulding09}. Gre10: \citet{greene2010}.}
\label{table:sou_info}
\end{table*}

\begin{table*}
\caption{\bf {\it XMM-Newton} Data Observation Details}
\small{
\begin{center}
\begin{tabular}{lccc}
\hline
\hline
\multicolumn{1}{c}{Name} &
\multicolumn{1}{c}{Obs. Date} &
\multicolumn{1}{c}{Exposure (pn/MOS1/MOS2)} &
\multicolumn{1}{c}{Filter (pn/MOS1/MOS2) } \\
\multicolumn{1}{c}{(1)} &
\multicolumn{1}{c}{(2)} &
\multicolumn{1}{c}{(3)} &
\multicolumn{1}{c}{(4)} \\
\hline
\hline
NGC~613  & 2010-12-30 & 32\,120/40\,810/40\,770 & medium/medium/medium \\
VIIZw73  & 2010-03-04 & 10\,080/14\,330/14\,600 & thin/thin/thin  \\
UGC~3789 & 2010-09-12 & 20\,900/26\,860/26\,880 & thin/thin/thin  \\
IRAS~16288+3929 & 2011-09-08 &  18\,568/21\,096/21\,098 & thin/thin/thin\\
NGC~6264 & 2010-08-26 & 45\,700/52\,270/52\,350 & thin/thin/thin  \\
\hline
\end{tabular}
\end{center}
Notes: (1): Galaxy name. (2): Observation date.
(3): pn, MOS1, and MOS2 observation exposures in sec. (4):
pn, MOS1, and MOS2 filters.} 
\label{table:obs_info}
\end{table*}

\section{The {\it XMM-Newton} spectral analysis}\label{sect:analysis}
 
\begin{figure*}
\includegraphics[width=\textwidth]{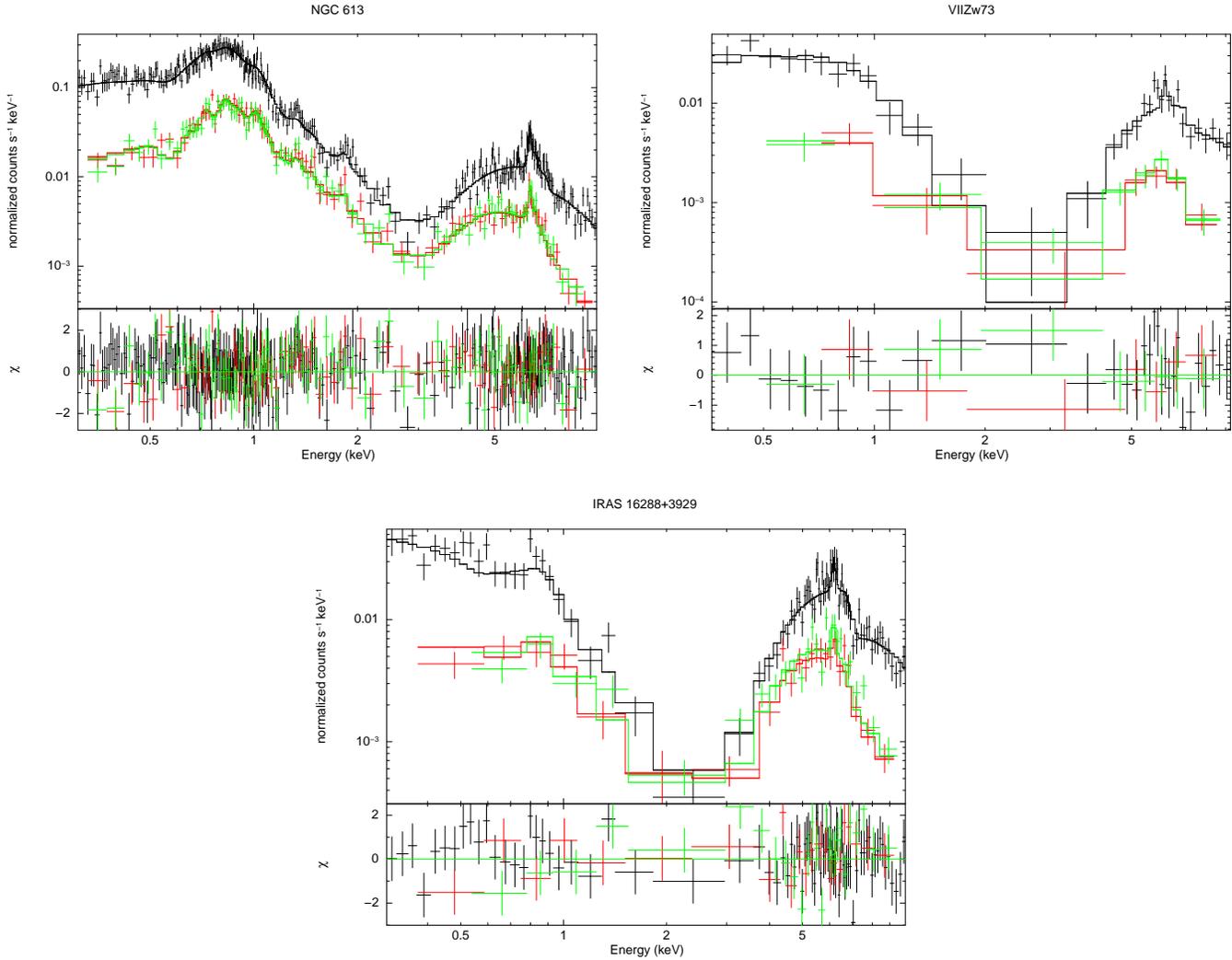}
\caption{{\it XMM-Newton} spectra, best fit models and residuals for NGC\,613, VIIZw73, and IRAS~16288+3929. EPIC pn data are shown in black, while MOS~1 and MOS~2 data are displayed in red and in green, respectively.}
\label{fig:nondisk}
\end{figure*}

\begin{figure*}
\includegraphics[width=\textwidth]{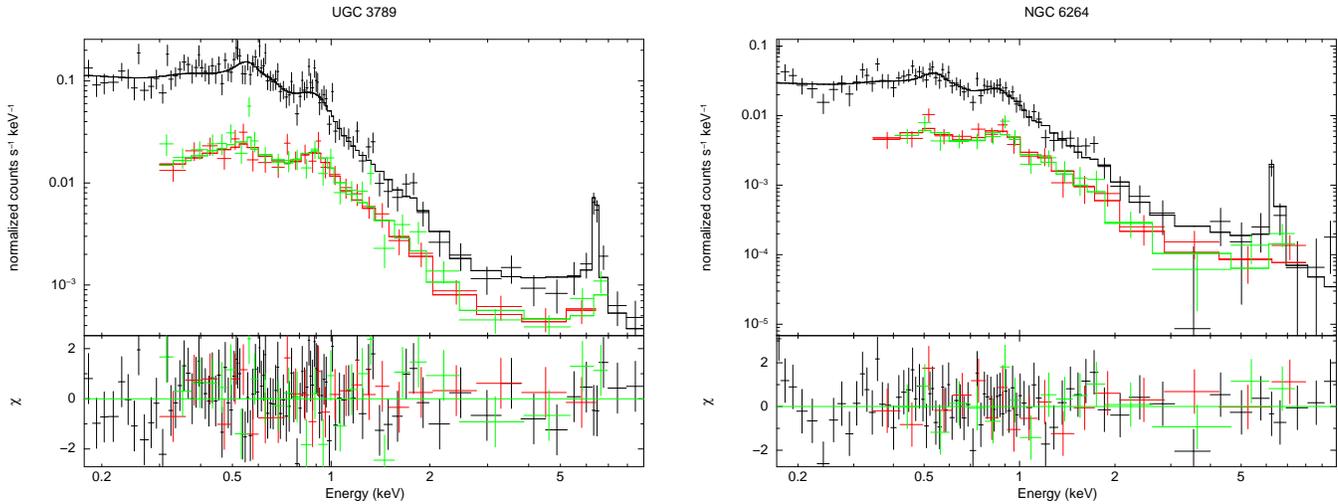}
\caption{{\it XMM-Newton} spectra, best fit models and residuals for the two disk-maser galaxies, UGC\,3789 and NGC~6264. EPIC pn data are shown in black, while MOS~1 and MOS~2 data are displayed in red and in green, respectively.}
\label{fig:disk}
\end{figure*}

From a first inspection of the spectra of our sources, they all show prominent soft ($<$2\,keV) X-ray emission (Fig.~\ref{fig:nondisk} and Fig.~\ref{fig:disk}). This ``soft excess'' is a characteristic of obscured AGN and its origin is still a matter of debate (see discussion in Sect.~\ref{sect:soft_em}). The limited energy resolution of the EPIC camera and the modest photon statistics in our observations, prevented us from accurately modelling the soft X-ray emission in our targets. Therefore, we used simplified models with one or two emission components from hot diffuse gas ({\sc mekal} in Xspec) and/or a soft power-law to describe the spectra below 2\,keV. Three soft components are significantly required to fit the spectra of NGC~613, UGC~3789 and NGC~6264, while the soft X-ray emission in IRAS~16288+3929 is well described by a soft power-law plus a thermal plasma component (Table~\ref{table:soft_par}). The {\it XMM-Newton} observation of VIIZw73 is heavily affected by solar flares reducing significantly the effective exposure (from 33\,ks down to $\sim$ 10 ks). The resulting spectrum is of poor statistics and we modeled the soft X-ray emission in this source with a simple black body component. 

Although all the spectra show a clear emission feature around 6 keV, there is a notable difference between the hard ($> $2\,keV) X-ray continuum of the two disk maser galaxies, UGC~3789 and NGC~6264, and the others. Indeed, the hard X-ray continuum appears to be almost completely suppressed in the former galaxies (Fig.~\ref{fig:disk}), while in the latter the primary emission of the AGN seems to be still directly visible (Fig.~\ref{fig:nondisk}). Accordingly, the best fit models for NGC~613, VIIZw73, and IRAS~16288+3929, have been obtained by adding to the soft components mentioned above, an absorbed power-law plus a narrow Gaussian component to account for the Fe K$\alpha$ emission around 6.4\,keV. In NGC~613 and VIIZw73, residuals are still visible suggesting the presence of an extra emission line. The addition of a second Gaussian at $6.61\pm 0.05$\,keV, significantly (at a probability of 99.6\%) improves the fit in the case of NGC~613, suggesting the presence of highly ionized material possibly related to the ionized soft spectrum. In VIIZw73, the Gaussian profile at $\sim$6.4\,keV is significant at only $\sim$ 98\% probability and there is evidence for the presence of a second emission feature at lower energies ($\sim$6.1\,keV). However, the poor statistics do not permit to properly characterize its line parameters. An absorption edge at $\sim$7\,keV, with an optical depth of 0.4$\pm$0.2 is instead required to fit the spectrum of IRAS~16288+3929. In the final model for this source, we fixed the edge energy to the best fit value (7.0$^{+0.1}_{-0.2}$\,keV, consistent with the absorption energy of neutral iron at 7.11\,keV) to better constrain the parameters we are most interested in (the absorbing column density, $N_{\rm H}$, and the photon index of the power law, $\Gamma$). For the three galaxies NGC~613, VIIZw73, and IRAS~16288+3929 , we measured intrinsic column densities in the range 4--5$\times 10^{23}$\,cm$^{-2}$. Consistently, the equivalent widths of the Fe~K$\alpha$ lines are $<$300\,eV, indicating that the transmitted component is heavily obscured but not completely suppressed (see discussion in Sect~\ref{sect:diagrams}). 

The hard X-ray emission of the two disk-maser galaxies, UGC~3789 and NGC~6264, instead, can be well described by two formally-acceptable models: a ``transmission model'', as the one illustrated above, consisting of an absorbed power-law plus a narrow Gaussian component and a ``reflection model'' in which the absorbed power-law is replaced with a reflection component \citep[the {\sc pexrav} model in Xspec;][]{magdziarz95}. In both scenarios we fixed the photon index of the power law to 1.8, a typical value for radio quiet AGN \citep[e.\,g.][]{piconcelli05}. In the case of the ``transmission model'', we estimated intrinsic column densities of 3.5$\times 10^{23}$\,cm$^{-2}$ and $>$4.6$\times 10^{23}$\,cm$^{-2}$ for UGC~3789 and NGC~6264, respectively. The equivalent width of the Fe K$\alpha$ line is 1.40$^{+0.90}_{-0.45}$\,keV for UGC~3789 and a lower limit of 930\,eV is obtained for NGC~6264. These values suggest the presence of Compton-thick absorbers along the line of sight (see Sect~\ref{sect:diagrams}). In the ``reflection scenario'', we fixed the cut-off energy and the inclination angle of the torus to the default values (100\,keV and $\sim$60{\degr}, respectively). Using the disk inclination angle inferred from the H$_2$O maser observations ($\sim$90{\degr}; \citealt{kuo2013,reid2013}), would not lead to significant changes. In addition, the maser disk and the torus might not be perfectly aligned \citep*[e.\,g.][]{gallimore04}. Therefore, we prefer to use the default value. We also assume that the continuum is entirely reflected (no direct component). The measured equivalent widths of the iron lines are 1.58$^{+0.44}_{-0.43}$\,keV and 3.72$^{+1.32}_{-1.31}$\,keV for UGC~3789 and NGC~6264, respectively. Although both scenarios give statistically acceptable fits, the detection of strong ($EW >$1\,keV) Fe K$\alpha$ lines indicates that the spectra of these sources are most likely due to pure reflection from cold material (see discussion in Sect~\ref{sect:diagrams}). 

The fluxes in the 0.1--2 and 2--10\,keV bands obtained by the spectral fitting of {\it XMM-Newton} data (Table~\ref{table:soft_par} and \ref{table:hard_par}) are consistent (within the errors) with those estimated from the analysis of Swift-XRT spectra (Castangia et al. in prep.). Galactic absorption (as in Table~\ref{table:soft_par}) has been added to our models. Initially, we allowed the column density to vary. However, since the best fitting values were found to be lower or equal than the Galactic ones, we fixed $N_{\rm H}$ to the Galactic values, which were taken from the LAB \citep[Leiden-Argentine-Bonn Galactic H{\sc i} Survey;][]{kalberla05} database. The resulting best fit spectral parameters are shown in Tables~\ref{table:soft_par} and \ref{table:hard_par}. Errors given in the table are calculated at 90\% confidence for one interesting parameter ($\Delta \chi ^2 = 2.71$).
 
\begin{table*}
\footnotesize{
\caption{\bf {\it XMM-Newton} spectral analysis. Soft component}
\label{table:soft_par}
\begin{center}
\begin{tabular}{llcccccccccr}
\hline
\hline
\multicolumn{1}{c}{Name} &
\multicolumn{1}{c}{Model} &
\multicolumn{1}{c}{$N_{\rm H, gal}$} &
\multicolumn{1}{c}{kT$_1$} &
\multicolumn{1}{c}{kT$_2$} &
\multicolumn{1}{c}{$\Gamma_{\rm soft}$} &
\multicolumn{1}{c}{F$_{\rm 0.1-2}$} &
\multicolumn{1}{c}{L$_{\rm 0.1-2}$} &
\multicolumn{1}{c}{L$_{\rm 0.1-2, corr}$} &
\multicolumn{1}{c}{$\chi^{2}$/dof} \\
\multicolumn{1}{c}{(1)} &
\multicolumn{1}{c}{(2)} &
\multicolumn{1}{c}{(3)} &
\multicolumn{1}{c}{(4)} &
\multicolumn{1}{c}{(5)} &
\multicolumn{1}{c}{(6)} &
\multicolumn{1}{c}{(7)} &
\multicolumn{1}{c}{(8)} &
\multicolumn{1}{c}{(9)} \\
\hline
\hline
NGC~613  & T$^*$ & 1.62 & 0.55$^{+0.03}_{-0.01}$ & 1.10$^{+0.13}_{-0.08}$ & 2.13$^{+0.04}_{-0.09}$ & 2.9 & 40.2 & 41.3 & 470/454 \\
VIIZw73  & T$^*$ & 6.76 & 0.16$^{+0.02}_{-0.01}$ &  $-$                  &  $-$                 & 0.3 & 41.1 & 42.6 & 32/39   \\
UGC~3789 & T & 5.2  & 0.16$\pm$0.02        & 0.71$^{+0.07}_{-0.06}$ & 2.8$\pm0.1$           & 1.4 & 40.6 & 41.4 & 171/145 \\
         & R &      & 0.17$^{+0.01}_{-0.02}$ & 0.71$^{+0.08}_{-0.05}$ & 3.0$^{+0.1}_{-0.2}$   & 1.4 & 40.6 &  41.2 & 168/146 \\
IRAS~16288+3929 & T$^*$ & 0.86 & --             & 0.7$\pm$0.1           & 3.1$\pm$0.3          & 0.6 & 41.1 & 42.5 & 145/125 \\
NGC~6264 & T & 5.32 & 0.14$\pm$0.01         & 0.72$^{+0.09}_{-0.06}$ & 2.7$\pm$0.2         & 0.4 & 41.0 & 41.8 & 88/99 \\
         & R &      & 0.15$^{+0.03}_{-0.02}$ & 0.63$\pm$0.04         & 2.8$^{+0.2}_{-0.3}$  & 0.4 & 41.0 & 41.6 & 92/100 \\
\hline				
\end{tabular}
\end{center}
Note: (1): Galaxy name.
(2): Adopted models. Model ``T'': transmission model {\sc (wabs(po+2mekal+wabs(po+zgauss))} in Xspec). 
Model ``R'': reflection model {\sc (wabs(po+2mekal+pexrav+zgauss)} in Xspec). 
``T$^*$'': in NGC\,613 a second Gaussian profile was included in the fit; 
the soft Xray emission in VIIZw73 was modeled with a simple black body;
in IRAS~16288+3929 only one thermal plasma was required and an absorption edge was included in the fit.     
(3): Galactic column density in units of 10$^{20}$ cm$^{-2}$. 
(4): Temperature of 1st {\sc mekal} in keV. 
(5): Temperature of the 2nd {\sc mekal} in keV. 
(6): Photon-index. 
(7): Model observed fluxes in the 0.1--2 keV band in units of 10$^{-13}$ erg cm$^{-2}$ s$^{-1}$. 
(8): Logarithm of the luminosity in the 0.1--2 keV band. 
(9): Logarithm of the luminosity in the 0.1--2 keV band corrected for galactic and intrinsic absorption (Table~\ref{table:hard_par}). 
(10): Chi-squared and degrees of freedom.}
\end{table*}


\begin{table*}
\footnotesize{
\caption{\bf {\it XMM-Newton} spectral analysis. Hard component}
\label{table:hard_par}
\begin{center}
\begin{tabular}{llcccccccr}
\hline
\hline
\multicolumn{1}{c}{Name} &
\multicolumn{1}{c}{Model} &
\multicolumn{1}{c}{$N_{\rm H, int}$} &
\multicolumn{1}{c}{$\Gamma$} &
\multicolumn{1}{c}{E} &
\multicolumn{1}{c}{EW} &
\multicolumn{1}{c}{F$_{\rm 2-10}$} &
\multicolumn{1}{c}{L$_{\rm 2-10}$} &
\multicolumn{1}{c}{L$_{\rm 2-10, corr}$} &
\multicolumn{1}{c}{$\chi^{2}$/dof} \\
\multicolumn{1}{c}{(1)} &
\multicolumn{1}{c}{(2)} &
\multicolumn{1}{c}{(3)} &
\multicolumn{1}{c}{(4)} &
\multicolumn{1}{c}{(5)} &
\multicolumn{1}{c}{(6)} &
\multicolumn{1}{c}{(7)} &
\multicolumn{1}{c}{(8)} &
\multicolumn{1}{c}{(9)} \\
\hline
\hline
NGC~613  & T$^*$ & 36$^{+5}_{-4}$   & 1.76$\pm$0.06     & 6.36$\pm$0.02  & 0.25$\pm$0.05 & 10.7  & 40.8 & 41.3  & 470/454 \\
         &   &                 &                    & 6.61$\pm$0.05 & 0.08$\pm$0.04 &       &      &        &         \\
VIIZw73  & T$^*$ & 49$^{+10}_{-15}$ & 1.2$^{+0.2}_{-0.6}$ & 6.43$^{+0.08}_{-0.09}$ & $<$0.31 & 7.4  & 42.4 & 43.0 & 32/39  \\
UGC~3789 & T & 35$^{+29}_{-17}$ & 1.8$^{f}$  & 6.45$\pm$0.04  & 1.40$^{+0.90}_{-0.45}$ & 1.4   & 40.6 & 41.0 & 171/145 \\
         & R & $-$             & 1.8$^{f}$  & 6.45$\pm$0.04  & 1.58$^{+0.44}_{-0.43}$ & 1.3   & 40.6 & $-$   & 168/146 \\
IRAS~16288+3929 & T$^*$ & 40$^{+5}_{-7}$ & 1.2$^{+0.3}_{-0.4}$ & 6.36$\pm$0.04 & 0.15$\pm$0.05 & 12.5 & 42.4 & 42.9 & 145/125 \\
NGC~6264  & T & $>$46  & 1.8$^{f}$  & 6.50$^{+0.05}_{-0.10}$ & $>$0.93               & 0.3    & 40.9 & 41.4 & 88/99 \\
          & R & $-$    & 1.8$^{f}$  & 6.52$^{+0.04}_{-0.07}$ & 3.72$^{+1.32}_{-1.31}$ & 0.3    & 40.8 & $-$   & 92/100 \\
\hline				
\end{tabular}
\end{center}
Note: (1): Galaxy name.
(2): Adopted models. Model ``T'': transmission model {\sc (wabs(po+2mekal+wabs(po+zgauss))} in Xspec). 
Model ``R'': reflection model {\sc (wabs(po+2mekal+pexrav+zgauss)} in Xspec). 
``T$^*$'': in NGC\,613 a second Gaussian profile was included in the fit; 
the soft Xray emission in VIIZw73 was modeled with a simple black body;
in IRAS~16288+3929 only one thermal plasma was required and an absorption edge was included in the fit.
(3): Intrinsic column density in units of 10$^{22}$ cm$^{-2}$. 
(4) Power-law photon index. 
(5): Rest energy of the line in keV. 
(6): Equivalent width of the line in keV. 
(7): Model observed fluxes in the 2-10 keV band in units of 10$^{-13}$ erg cm$^{-2}$ s$^{-1}$.
(8): Logarithm of the luminosity in the 2-10 keV band.
(9): Logarithm of the luminosity in the 2-10 keV band corrected for intrinsic absorption.
(10): Chi-squared and degrees of freedom.
}
\end{table*}


\section{Broad band spectral fitting of VIIZw73 and IRAS~16288+3929}
\begin{figure*}
\includegraphics[width=\textwidth]{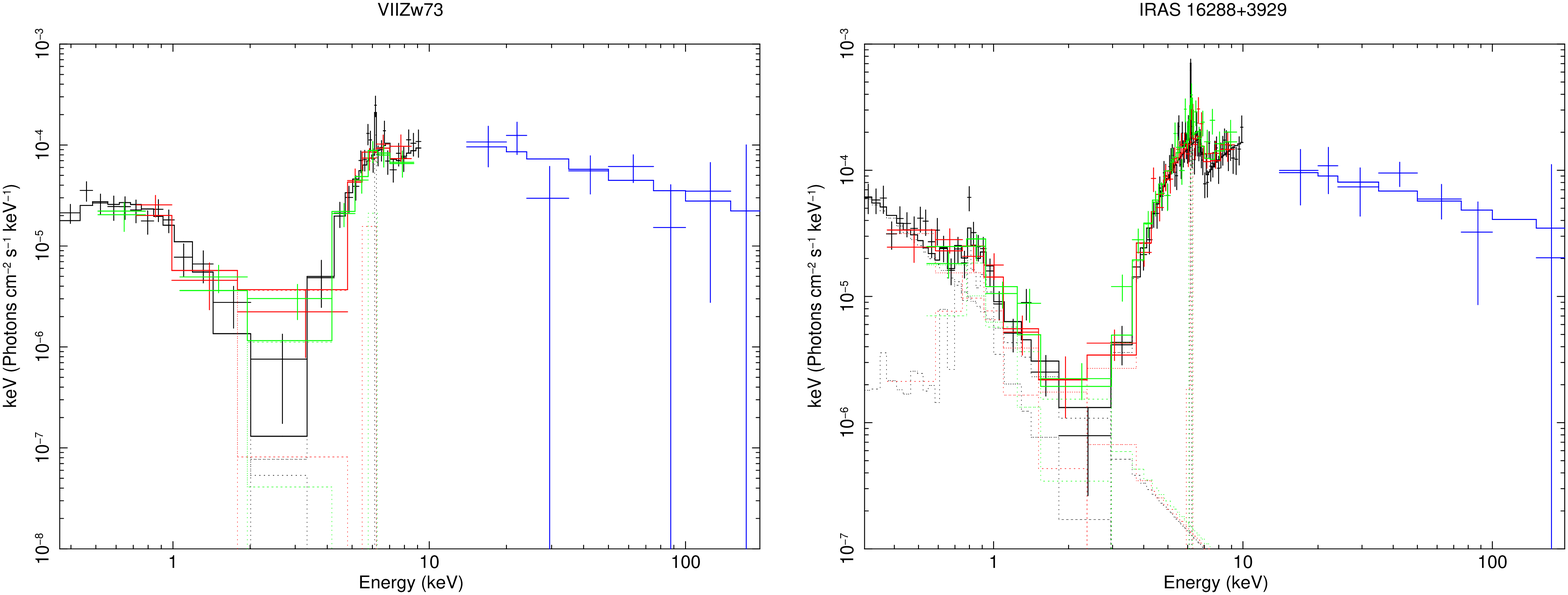}
\caption{Broad band ({\it XMM-Newton+Swift-BAT}) X-ray spectra, best fit models and residuals for VIIZw73 and IRAS~16288+3929. EPIC pn data are shown in black, MOS~1 and MOS~2 data are displayed in red and in green, respectively, and BAT data are represented in blue.}
\label{fig:xmm+bat}
\end{figure*}

Both VIIZw73 and IRAS~16288+3929 were detected by {\it Swift-BAT} in the 70-Month Hard X-ray Survey with signal-to-noise-ratios of $\sim$6 \citep{baumgartner2013}. The spectra of the survey are available online and we have retrieved them together with the appropriate response files from the survey website\footnote{http://heasarc.gsfc.nasa.gov/docs/swift/results/bs70mon/}.
We fit {\it XMM-Newton} and {\it Swift-BAT} spectra simultaneously fixing the pn normalization to 1 and leaving the MOS1/2 and BAT normalizations free to vary (Fig.~\ref{fig:xmm+bat}). We find that our best fit models to the $<$10\,keV data can account also for the high energy emission ($\chi^2$/dof=37/46 and 157/131 for VIIZw73 and IRAS~16288+3929, respectively; see also Fig.~\ref{fig:xmm+bat}). In the model of IRAS~16288+3929 we let the energy of the absorption edge free to vary. We find that the energy is 7.0$^{+0.1}_{-0.2}$\,keV, in agreement with the value derived from the {\it XMM-Newton} spectral analysis (Sect.~\ref{sect:analysis}). We note that, although the estimated model parameters coincide, within the uncertainties, with those reported in Tables~\ref{table:soft_par} and \ref{table:hard_par}, the power-law representing the primary AGN emission has a steeper spectral index ($\Gamma=1.7^{+0.1}_{-0.3}$ and 1.5$\pm$0.3) and is absorbed by a larger column density ($56^{+10}_{-12}$ and $43\pm5 \times 10^{22}$\,cm$^{-2}$, for VIIZw73 and IRAS~16288+3929, respectively) when fitting the entire broad-band spectrum instead of the {\it XMM-Newton} data alone. The statistics of the BAT spectra do not allow us to constrain the high-energy cut-off value.
VIIZw73 has an intrinsic luminosity of 1.41$\times 10^{43}$\,erg\,s$^{-1}$ (3.76$\times 10^{43}$\,erg\,s$^{-1}$) in the 2--10\,keV (15--150\,keV) band. The inferred intrinsic 2--10\,keV (15--150\,keV) luminosity of IRAS~16288+3929 is 1.01$\times 10^{43}$\,erg\,s$^{-1}$ (2.54$\times 10^{43}$\,erg\,s$^{-1}$). 

\section{Discussion}

\subsection{Soft X-ray emission}\label{sect:soft_em}
The soft X-ray emission in obscured AGN could be produced either by a collisionally ionized gas (heated by shocks induced by AGN outflows or by starburst powered galactic winds) or by the gas photoionized by the AGN primary emission \citep[][and references therein]{guainazzi07}. Recently, however, high resolution spectra of 69 obscured AGN obtained with gratings onboard {\it Chandra} and {\it XMM-Newton} revealed that this ``soft excess'' is due to the blending of a wealth of emission lines (mainly from highly ionized species and L-shell transitions of Fe), with a very low level of continuum emission, which are produced in a photoionized gas \citep{guainazzi07}. The active nucleus is believed to be the dominant source of photoionization, although star formation may also contribute to the process \citep[e.\,g.][]{bianchi07}. In order to assess whether a significant contribution to the soft X-ray emission belongs to star formation, we calculated the star formation rates, {\it SFR}, for the target galaxies, using the \citet{kennicutt98} relation:
$$
SFR=\frac{L_{\rm FIR}}{5.9\times10^9L_{\odot}}M_{\odot}{\rm yr^{-1}},
$$
where the far--infrared (FIR) luminosity, $L_{\rm FIR}$, is determined from the combination of the {\it IRAS} 60\,$\mu$m and 100\,$\mu$m fluxes in Table~\ref{table:sou_info}, according to the formula reported in \citet*{helou85}. The FIR fluxes and the calculated {\it SFR} are listed in Table~\ref{table:sfr}. In our estimates, we have assumed that the FIR luminosity is entirely produced by the star formation process, neglecting the possible contribution of the AGN (torus). Therefore, the estimated {\it SFR} are upper limits to the true values. Furthermore, because of the coarse angular resolution of {\it IRAS}, the FIR fluxes are integrated over the entire galaxies, hence, the {\it SFR} calculated in this way is the galactic and not the nuclear {\it SFR}. Since NGC\,6264 lacks FIR data, we calculated the star formation rate for this galaxy from the 1.4\,GHz radio flux density through the formula \citep{condon92}:
$$
SFR=\frac{L_{\rm 1.4GHz}}{4\times10^{28}{\rm erg s^{-1} Hz^{-1}}}M_{\odot}{\rm yr^{-1}}.
$$  
The 1.4\,GHz flux density was taken from the NVSS survey and is 2.3\,mJy.
With the exception of VIIZw73, all our targets have star formation rates similar to our Galaxy ($\sim$2\,M$_{\odot}$yr$^{-1}$; \citealt{kennicutt2012}). Therefore, it is unlikely that star formation can contribute significantly to the soft X-ray emission. The {\it SFR} of $\sim$13\,M$_{\odot}$\,yr$^{-1}$ estimated for VIIZw73 is typical of starburst galaxies like M\,82 and may dominate the spectrum below 2 keV through the thermal emission of supernova powered winds \citep{persic04}. Therefore, the starburst contribution to the soft X-ray emission may be non negligible in the case of VIIZw73.  Profiting from the soft X-ray/FIR and soft X-ray/radio correlations reported in \citet*{ranalli03} (Eqs. 8 and 9), we have also derived the expected starburst luminosity in the 0.5--2\,keV energy range, for all the targets. Comparing the expected with the observed luminosities (corrected for Galactic and intrinsic absorption) we find that star formation contributes between 5\% and 20\% to the observed soft X-ray emission, with the exception of VIIZw73 in which the starburst contribution is of $\sim$40\%. These percentages, however, should be considered as upper limits. In fact, the aforementioned correlations are derived for normal or starburst galaxies were the FIR luminosity is entirely produced by star formation, while a non negligible fraction of the FIR luminosity in our targets is likely to be produced by the AGN.   


 \begin{table*}
\footnotesize{
\caption{\bf Estimated $L_{\rm FIR}$ and $SFR$ for the target galaxies.}
\label{table:sfr}
\begin{center}
\begin{tabular}{lcccccr}
\hline
\hline
\multicolumn{1}{c}{Name} &
\multicolumn{1}{c}{$L_{\rm FIR}$} &
\multicolumn{1}{c}{$SFR$} &
\multicolumn{1}{c}{$F_{2-10}/F_{[\rm OIII]}$} &
\multicolumn{1}{c}{$\log(L_{2-10}/L_{[\rm FIR]})$} &
\multicolumn{1}{c}{$F_{2-12}/F_{25} \nu_{25}$} &
\multicolumn{1}{c}{HR4} \\
\multicolumn{1}{c}{(1)} &
\multicolumn{1}{c}{(2)} &
\multicolumn{1}{c}{(3)} &
\multicolumn{1}{c}{(4)} &
\multicolumn{1}{c}{(5)} &
\multicolumn{1}{c}{(6)} &
\multicolumn{1}{c}{(7)} \\
\hline
\hline
NGC~613  & 1.5 & 2.7  & 5.6 & -3.3 & 0.006 & 0.6 \\
VIIZw73  & 7.8 & 13.4 & 10  & -2.5 & 0.019 & 0.8 \\
UGC~3789 & 0.6 & 1.1  & 0.03 & -3.2 & 0.004 & 0.4 \\
IRAS~16288+3929 & 1.4 & 2.4 & 7.3 & -1.7 & 0.080 & 0.8 \\
NGC~6264  & -- & 1.4  & 0.2 & --   &  --   & --    \\
\hline				
\end{tabular}
\end{center}
Note: (1): Galaxy name. (2): FIR luminosity in units of 10$^{10}$\,L$_{\odot}$, calculated from the IRAS 60 and 100\,$\mu$m fluxes using the formula reported in \citet{helou85}. (3): Star formation rate ($SFR$) in units of M$_\odot$\,yr$^{-1}$. (4): Ratio between the 2--10\,keV flux and the [O{\sc iii}] line flux corrected for Galactic and intrinsic extinction (see Sect.~\ref{sect:diagrams}. (5): Logarithm of the ratio between the observed 2--10\,keV luminosity and the FIR luminosity, where $L_{\rm FIR}$ is calculated using fluxes from all the four IRAS bands (12, 25, 60 and 100\,$\mu$m) following the method reported in \citet{zhang2010}. (6): Ratio between the 2--12\,keV flux and the IRAS mid-infrared flux at 25\,$\mu$m. (7): {\it XMM-Newton} color HR4, defined using the 2--4.5 and 4.5--12\,keV bands.
}
\end{table*}

\subsection{H$_2$O masers and nuclear obscuration}\label{sect:diagrams}
Spectral analysis of the observed maser galaxies reveals that all of them host heavily obscured or Compton-thick AGN. In three objects, NGC\,613, VIIZw73, and IRAS~16288+3929, we have been able to estimate the amount of intrinsic absorption, measuring column densities of 4--6$\times$10$^{23}$\,cm$^{-2}$ (Table~\ref{table:hard_par}). In the case of the two disk-maser galaxies, UGC\,3789 and NGC~6264, the primary continuum is strongly suppressed in the 2--10\,keV energy range and the spectral analysis does not allow us to distinguish between a transmission or a pure reflection scenario nor to determine the column density (Sect.~\ref{sect:analysis}). The presence of Compton-thick material along the line of sight may be inferred from the large equivalent widths ($EW>$1\,keV; Table~\ref{table:hard_par}) of the iron 6.4\,keV lines. In fact, the fluorescent Fe K$\alpha$ line is thought to arise from the torus with $EW <$200\,eV but absorbing column densities $>$10$^{23}$\,cm$^{-2}$, reducing the continuum underneath the iron line, may increase the equivalent width up to values of a few keV \citep[e.\,g.][]{maiolino98,murphy09}. $EW$(Fe K$\alpha$)$>$1\,keV indicates complete absorption of the AGN primary continuum and is used to identify Compton-thick AGN \citep[e.\,g.][]{maiolino98}. 

In order to confirm the Compton-thin or -thick nature of our targets, we have also considered the ratio $T$ between the 2--10\,keV X-ray flux and the [O{\sc iii}] $\lambda$5007 flux (Table~\ref{table:sfr}). Indeed, as shown in \citet{bassani99}, a flux ratio $T \leq 1$ is indicative of absorbing column densities in excess of 10$^{24}$\,cm$^{-2}$. After correcting the observed [O{\sc iii}] flux (Table~\ref{table:sou_info}) for Galactic and intrinsic extinction following the method in \citet{bassani99}, we derive the ratio $T=F_{\rm 2-10 keV}/F_{[\rm OIII]}$ for all our targets. We obtain $T$=5.6, 10 and 7.3 for NGC\,613, VIIZw73, and IRAS~16288+3929, respectively. These values are consistent with their Compton-thin nature. For UGC~3789 and NGC~6264, instead, we calculate  $T=0.03$ and $T=0.2$, respectively. These ratios, together with the estimated equivalent widths of the iron K$\alpha$ lines, places these objects among the Compton-thick AGN in the diagram of \citet{bassani99}. Another isotropic indicator of the intrinsic brightness of the AGN is the far-infrared (FIR) emission. Therefore we have also estimated the position of our target in the diagnostic diagram of \citet{zhang2010} and the very recent one reported in \citet*{severgnini2012}. The former, using archival data for 76 H$_2$O maser galaxies, were able to identify a region in the $\log(L_{2-10}/L_{[\rm FIR]})$ vs. $\log EW({\rm K \alpha})$ plane that is populated by Compton-thick sources. This region is defined approximately by $\log(L_{2-10}/L_{[\rm FIR]}) < -2.75$ and $\log EW({\rm K \alpha})> 2.5$. \citet{severgnini2012} instead, utilize the combination of the $F_{\rm 2-12 keV}/ \nu_{25} F_{25}$ ratio (where $F_{25}$ is the IRAS 25$\mu$m flux density) and the {\it XMM-Newton} color HR4, defined using the 2--4.5 and 4.5--12\,keV bands \citep{watson09}. Compton-thick AGN have $[F_{\rm 2-12 keV}/ \nu_{25} F_{25}] < 0.02$ and HR4$> -0.2$. As can be seen by looking at Table~\ref{table:sfr}, the position of UGC\,3789 falls into the quadrant of the Compton-thick AGN candidates in both diagnostic diagrams. These diagrams also confirm the Compton-thin nature of VIIZw73 and IRAS~16288+3929. Surprisingly, NGC\,613 is Compton-thin according to \cite{zhang2010} while its position in the $[F_{\rm 2-12 keV}/ \nu_{25} F_{25}]$ vs. HR4 plane is that of Compton-thick sources. NGC\,6264, unfortunately, lacks infrared data. To summarize, combining the results of the spectral analysis with the indication of the diagnostic plots, we find that NGC\,613, VIIZw73, and IRAS~16288+3929 are heavily absorbed but Compton-thin, while intrinsic absorption in excess of 10$^{24}$\,cm$^{-2}$ is likely to be present in UGC\,3789 and NGC~6264. In the following we will consider UGC\,3789 and NGC~6264 as ``bona fide'' Compton-thick AGN although hard X-ray data at energies $>$10\,keV would be necessary to conclusively determine their nature.    

Adding the five galaxies studied in this paper to the sample of 42 water maser sources examined by \citet{greenhill08} (see Sect.~\ref{sect:intro}), we obtain a final sample of 42+5=47 objects, 96\% of which (45/47) show $N_{\rm H} >$10$^{23}$\,cm$^{-2}$ and 57\% (27/47) have $N_{\rm H} >$10$^{24}$\,cm$^{-2}$. The percentange of disk-maser galaxies hosting Compton-thick AGN is found to be 78\% with 18/23 sources with $N_{\rm H} >$10$^{24}$\,cm$^{-2}$. The five Compton-thin disk-masers in the sample are Mrk\,1210, NGC\,4051, NGC\,4258, NGC\,4388, and 3C\,403 (for references to the $N_{\rm H}$ estimates of NGC\,4258 and NGC\,4388, see Table~\ref{table:diskmasers}; for the other sources see \citealt{zhang06} and references therein). 
However, NGC\,4051 and 3C\,403 might not be disk-masers. Indeed, despite being included in the disk-maser sub-sample by \citet{greenhill08}, the origin of the maser emission in these galaxies is still under debate, with the jet/outflow association being the most likely option \citep{tarchi07,tarchi2011b}. If we exclude NGC\,4051 and 3C\,403 from the sub-sample of disk-masers, the fraction of Compton-thick AGN becomes 18/21 (86\%). Interestingly, the two remaining Compton-thin disk-masers, NGC\,4258 and NGC\,4388, are known to be strongly variable X-ray sources, with reported $N_{\rm H}$ changes on timescales of about one day or less \citep{yamada09,risaliti2010}. Although the measured $\Delta N_{\rm H}$ in the aforementioned galaxies is not so extreme,  X-ray absorption variability might sometimes lead to radical spectral variations, as in the case of NGC\,1365, that was found to change from Compton-thin to reflection dominated Compton-thick \citep[][and references therein]{risaliti2010}. Therefore, the percentage of Compton-thick disk-masers may potentially become even higher. Our results reinforce those obtained by \citet{greenhill08}, supporting the trend of water maser sources associated with AGN to show high column densities and, in particular, the strong connection between H$_2$O masers in accretion disks and Compton-thick AGN.

\subsection{Correlations between maser and X-ray derived properties of disk-maser galaxies}\label{sect:corr}
In case of accretion disks, maser and X-ray emission are also linked by the maser excitation mechanism. The standard model postulates that the accretion disk, at scales of 0.1--1\,pc, is heated by X-ray irradiation to temperatures of the order 300--1000\,K, suitable for 22\,GHz maser emission \citep*{neufeld94}. The model predicts a statistical increase of the inner radius of the masing disks as a function of X-ray luminosity. Within the framework of the model, a lower luminosity permits the survival of molecules (and hence of maser action) at smaller radii. Studying a small sample (8 galaxies) of disk maser hosts, \citet{tilak08} found that not only the inner radius but also the degree of clumpyness of the disk may increase with the intrinsic X-ray luminosity. With the aim of confirming this trend on a larger sample, we compiled in Table~\ref{table:diskmasers} an updated list of all disk-masers with precise (obtained with VLBI) measurements of the accretion disk radius, $R_{\rm H2O}$, and the black hole mass, $M_{\rm BH}$, for which the intrinsic X-ray luminosity in the 2--10\,keV band, $L^{\rm I}_{2-10}$, can be estimated. In addition to the objects analysed by \citet{tilak08}, the list contains 6 more galaxies, including the two disk-masers studied in this paper, UGC\,3789 and NGC\,6264, for which estimates of $R_{\rm H2O}$ and $M_{\rm BH}$ have been recently reported \citep{kuo2011,kuo2013,reid2013}. Intrinsic X-ray luminosities listed in Table~\ref{table:diskmasers} have been taken from the literature when possible. To have an estimate of $L^{\rm I}_{2-10}$ for Compton-thick sources with only a lower limit of the column density or in case the intrinsic luminosity was not reported in the literature, we have applied a correction factor of 60 to their observed one, $L_{2-10}$, as suggested in \citet{panessa06}. The bolometric luminosity was then obtained multiplying $L^{\rm I}_{2-10}$ by a factor of 20. We used this value of the bolometric correction to be consistent with \citet{tilak08}, though we note that for some of the AGN in Table~\ref{table:diskmasers}, which have $L^{\rm I}_{2-10} / L_{\rm Edd} \sim 10^{-2}$, the correction may be larger \citep{vasudevan07}. 

\begin{figure}
\includegraphics[scale=0.5]{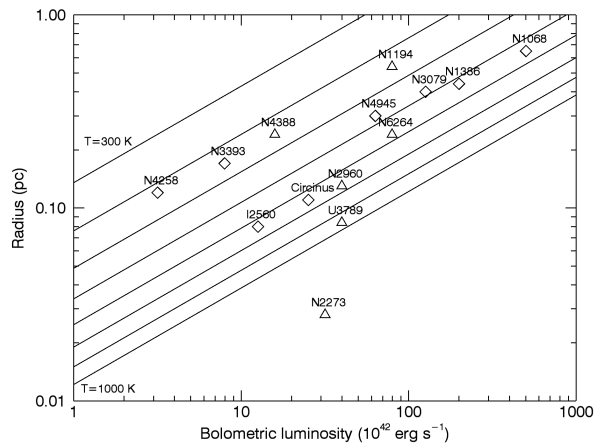}
\caption{Observed inner radii of the masing disks against AGN bolometric luminosities for the sources in Table \ref{table:diskmasers}. This figure is an updated version of Fig.~7 of \citet{tilak08}. The diamonds mark the position of the disk-masers already present in the plot of \citet{tilak08}, while the triangles are new sources. The labels indicate the galaxies' names. The solid lines represent the value of the disk radius predicted in the case of thermodynamic equilibrium ($R\sim \sqrt{L_{\rm Bol}/4 \pi \sigma T^4}$), for constant temperatures of 300, 400, 500, 600, 700, 800, 900, and 1000\,K from top to bottom.}
\label{fig:plot_avanti}
\end{figure}

In Fig.~\ref{fig:plot_avanti}, we present an updated version of Fig.~7 of \citet{tilak08}, where the inner radii of the masing disks are plotted against the AGN bolometric luminosities. If we assume, as a first approximation, that the disk is in thermodynamic equilibrium, than the radius is given by $R\sim \sqrt{L_{\rm Bol}/4 \pi \sigma T^4}$ and we can infer a rough estimate of the disk temperature \citep{tilak08}. Despite the large uncertainties in the bolometric luminosities, the observed inner radii of maser emission are consistent (except one case) with temperatures of 400-1000\,K (Fig.~\ref{fig:plot_avanti}), in agreement with the predictions of the model of \citet{neufeld94} for collisionally pumped water masers in X-ray irradiated circumnuclear gas. We fit a linear relationship to the data using the MPFITEXY routine \citep{markwardt09} and obtain a best fit temperature of $\sim$600\,K confirming the result previously reported by \citet{tilak08}.
In order to explain why the disk-masers with geometrically thick disks and flattened rotation curves (namely, NGC~1068, NGC~4945, NGC~3079, and NGC~1386), have the highest luminosity and largest radii, \citet{tilak08} argued that the higher the luminosity of the sources, the farther from the black hole the water emission arises and the less ordered and more clumpy is the disk. However, we do not confirm this hypothesis, at least for galaxies with luminosities lower than 10$^{44}$\,erg\,s$^{-1}$. For $L_{\rm Bol} < 10^{44}$\,erg\,s$^{-1}$ AGN with a comparable luminosity show accretion disks with different thickness/rotation curves. Indeed, the thin disk maser galaxies, NGC~1194 and NGC~6264, have a luminosity (7-8$\times 10^{43}$erg\,s$^{-1}$) similar to that of NGC~4945 that appears to hosts a thick accretion disk \citep*{greenhill97}.  

The disk in NGC~2273 appears to be different from the other 13 disk-masers we considered. The radius inferred from the water maser distribution is a factor $>3$ smaller than the radius predicted by our zeroth order analysis for the temperature range 400-1000\,K. The anomalous position of NGC~2273 in the $R_{\rm H2O}-L_{\rm Bol}$ plane might be due to the fact that we overestimated the bolometric luminosity (3.4$\times10^{43}$\,erg\,s$^{-1}$). Since the error associated to $L_{\rm Bol}$ is difficult to quantify, we have also estimated the bolometric luminosity through the absorption-corrected $L_{[\rm OIII]}$ \citep{greene2010}. Using the optical bolometric correction of \citet{lamastra09} we obtain a bolometric luminosity of $\sim$2.3$\times 10^{43}$\,erg\,s$^{-1}$ that is still too large to explain the small radius of the masing disk in NGC~2273. For a bolometric luminosity in the range 2.3--3.4$\times10^{43}$\,erg\,s$^{-1}$, the temperature of the disk at a distance of 0.028\,pc from the central engine should be (assuming thermodynamic equilibrium) 1450--1560\,K. According to the model of \citet{neufeld94}, at this temperature the gas should be atomic rather than molecular and no maser emission from H$_2$O should be produced. Indeed, when the interstellar medium is mostly atomic the gas-phase formation of H$_2$O is quenched by the low H$_2$ abundance. However, \citet{neufeld94} take into account only the formation of water via the gas-phase route, while recent models suggest that, in X-ray exposed environments, the formation of water on dust grains (which dominates over the gas-phase route when the gas is mostly atomic) might be very efficient \citep*{meijerink2012}. In order to inspect whether or not the dust may survive at 0.028\,pc from the black hole, we calculate the inner radius of the dust distribution in NGC~2273. Assuming a smooth dust distribution, for a typical dust sublimation temperature of $\sim$1500\,K we find $R_{\rm dust}$=0.06--0.07\,pc \citep{nenkova08b}. Therefore at the radius of the water maser emission, the dust grains are likely to be destroyed. The scenario is quite different if the medium is clumpy, because, in this case, the dust temperature is not uniform within the clump but, for a given radius, is higher on the illuminated face and lower on the dark face of the clump \citep{nenkova08a}. In particular, the dust temperature at a radius of 0.028\,pc varies between  2000--2150\,K and 550--590\,K within each clump \citep{nenkova08a}. If this model is correct, the formation of water on dust grains would then be possible in NGC~2273 even so close to the black hole.

\begin{figure}
\includegraphics[scale=0.5]{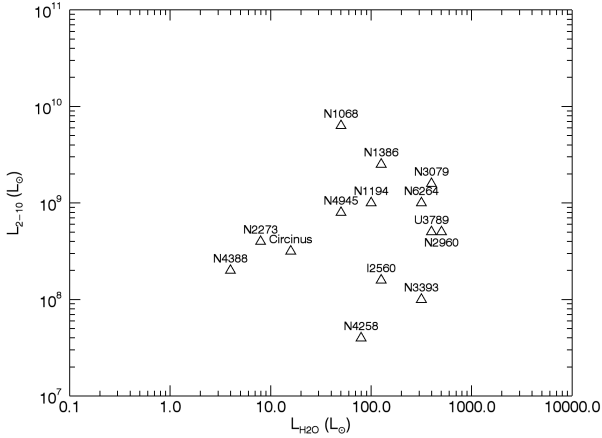}
\caption{Intrinsic (corrected for intrinsic absorption) X-ray luminosity in the 2--10\,keV band against total isotropic water maser luminosity for the sources in Table \ref{table:diskmasers}.} 
\label{fig:plot_lh2o_lx}
\end{figure}

A further possible relationship between the intrinsic X-ray luminosity in the 2--10\,keV band and the total isotropic maser luminosity has been reported by \citet{kondratko06a}, studying a sample of 30 water masers associated with AGN. They found that $L^{\rm I}_{2-10} \propto L_{\rm H2O}^{0.5}$, in agreement with the model of \citet{neufeld95}. This correlation suggests that water maser galaxies have similar AGN properties (Eddington ratios, bolometric corrections, viscosities, accretion efficiencies) and maser beaming angles. One might expect this correlation to be stronger for the sample of water masers associated with accretion disks as compared to the sample including masers of different origin, because of a more homogeneous geometry and because they are likely all produced by the same maser excitation mechanism. However, no correlation is apparent, according to \citet{kondratko06a}, considering the subset of disk-maser systems, probably due to the small size of the sample (10 objects). In Fig.~\ref{fig:plot_lh2o_lx} we plot $L^{\rm I}_{2-10}$ against $L_{\rm H2O}$ for the larger sample of disk-masers in Table~\ref{table:diskmasers}. 
A simple linear fit in the log-log plane, with the slope fixed to 0.5, yielded a $\chi^2$ of 13.2 for 13 d.o.f, hence, the data do not formally rule out (at the 0.05 significance) the predicted relation between X-ray and maser luminosity ($L^{\rm I}_{2-10} \propto L_{\rm H2O}^{0.5}$). However, given the large errors we considered (we took an average error of 0.5\,dex in both variables for all the sources) we treat this result with caution\footnote{If we calculate the Pearson correlation coefficient between  $L^{\rm I}_{2-10}$ and $L_{\rm H2O}$ we find that it is rather small ($r=0.14$), indicating that X-ray and H$_2$O luminosities are not correlated}. More disk-maser sources with good quality X-ray data and more precise estimates of  $L^{\rm I}_{2-10}$ and $L_{\rm H2O}$ are necessary to test the existence of a correlation between these two quantities.

\begin{table*}
\footnotesize{
\caption{\bf X-ray properties of confirmed disk-masers.}
\label{table:diskmasers}
\begin{center}
\begin{tabular}{lrcccccccc}
\hline
\hline
\multicolumn{1}{l}{Name} &
\multicolumn{1}{r}{Distance} &
\multicolumn{1}{c}{$N_{\rm H}$} &
\multicolumn{1}{c}{Ref.} &
\multicolumn{1}{c}{Log($L^{\rm I}_{2-10}$)} &
\multicolumn{1}{c}{Ref.} &
\multicolumn{1}{c}{log($M_{\rm BH}$)} &
\multicolumn{1}{c}{Disk size} &
\multicolumn{1}{c}{Ref.} &
\multicolumn{1}{c}{Log($L_{\rm H2O}$)} \\
\multicolumn{1}{c}{(1)} &
\multicolumn{1}{c}{(2)} &
\multicolumn{1}{c}{(3)} &
\multicolumn{1}{c}{(4)} &
\multicolumn{1}{c}{(5)} &
\multicolumn{1}{c}{(6)} &
\multicolumn{1}{c}{(7)} &
\multicolumn{1}{c}{(8)} &
\multicolumn{1}{c}{(9)} &
\multicolumn{1}{c}{(10)} \\

\hline
\hline
\textbf{NGC\,1068} & 16 & $\geq$100              & Mat97 & 43.4 & Til08 & 7.3  & 0.65--1.1  & Til08 & 1.7 \\
NGC\,1194          & 53 & 10.6$^{+3.6}_{-2.4}$   & Gre08 & 42.6 & Gre08 & 7.8  & 0.54--1.33 & Kuo11 & 2.0 \\
\textbf{NGC\,1386} & 12 & 20$^{+5}_{-3}$       & Fuk11 & 43.0 & Til08 & 6.1  & 0.44--0.94 & Til08 & 2.1 \\
NGC\,2273          & 26 & 15$\pm$4        & Awa09 & 42.2 & Awa09 & 6.9 & 0.028--0.084 & Ku011 & 0.9 \\
UGC\,3789          & 46 & $>$10              & this work & 42.3 & this work & 7.0 & 0.084--0.30 & Kuo11 & 2.6 \\
NGC\,2960          & 72 & $\geq$10               & Gre08 & 42.3 & Gre08 & 7.1  & 0.13--0.37    & Kuo11 & 2.7 \\
\textbf{NGC\,3079} & 15 & 54.0$^{+6.1}_{-6.5}$   & Bur11 & 42.8 & Til08 & 6.3  & $\sim$0.4     & Til08 & 2.6 \\
IC\,2560           & 40 & $\geq$10               & Til08 & 41.8 & Til08 & 6.5 & 0.08--0.27   & Gre09 & 2.1 \\
NGC\,3393          & 51 & 45.0$^{+6.2}_{-3.6}$   & Bur11 & 41.6 & Til08 & 7.5  & 0.17 & Til08  & 2.5 \\
NGC\,4258          & 7.2 & 0.87$\pm$0.03         & Cap06 & 41.2 & Til08 & 7.6  & 0.12--0.28    & Til08 & 1.9 \\
NGC\,4388          & 19 & 3.5$\pm$0.1         & Fuk11 & 41.9 & Cap06 & 6.9  & 0.24--0.29    & Kuo11 & 0.6 \\
\textbf{NGC\,4945} & 8  & 42.5$\pm$2.5           & Don03 & 42.5 & Til08 & 6.1 & $\sim$0.3     & Til08  & 1.7 \\
Circinus           & 6  & 43$^{+4}_{-7}$          & Mat99 & 42.1 & Til08 & 6.2  & 0.11--0.4     & Til08 & 1.2 \\
NGC\,6264          & 139 & $>$10             & this work & 42.6 & this work & 7.5 & 0.24--0.80  & Kuo11 & 2.5 \\
\hline				
\end{tabular}
\end{center}
Note: (1): Galaxy name. Thick disks with flattened rotation curves are in boldface. (2): Distance in Mpc. We adopted the distances used in the determination of the maser disk radii. (3): Intrinsic column density in units of 10$^{23}$\,cm$^{-2}$. (4): References for $N_{\rm H}$; Awa09: \citet{awaki09}; Bur11: \citet{burlon2011}; Cap06: \citet{cappi06}; Don03: \citet{done03}; Fuk11: \citet{fukazawa2011}; Gre08: \citet{greenhill08}; Mat97: \citet{matt97}; Mat99: \citet{matt99}; Til08: \citet{tilak08}. (5): Logarithm of the intrinsic luminosity in the 2-10\,keV band (in erg\,s$^{-1}$cm$^{-2}$). For NGC\,2273, UGC\,3789, NGC\,2690, and NGC\,6264 we used the observed fluxes or the apparent luminosities multiplied by a factor of 60 (see Sect.\ref{sect:corr}) to estimate the intrinsic luminosities (6): References for Log($L^{\rm I}_{2-10}$). Awa09: \citet{awaki09}; Cap06: \citet{cappi06}; Gre08: \citet{greenhill08}; Til08: \citet[][and references therein]{tilak08}. (7) Logarithm of the black hole mass (in M$_{\odot}$). (8) Disk size in pc. (9) References for columns 8-9. Kuo11: \citet{kuo2011}; Gre09: \citet{greenhill09}; Til08: \citet[][and references therein]{tilak08}. (10): Logarithm of the maser isotropic luminosity (in L$_{\odot}$}). The H$_2$O luminosities are taken from \citet[][and references therein]{kondratko06a} for all the galaxies, with exception of NGC\,1194 (whose luminosity was derived directly from the spectrum), UGC\,3789, NGC\,2690, and NGC\,6264 (whose data are reported in \citet{tarchi2011a}).
\end{table*}


\section{Summary}
We observed five H$_2$O maser galaxies, NGC\,613, VII\,Zw\,73, UGC\,3789, IRAS\,16288+3929, and NGC\,6264, with {\it XMM-Newton} in the 0.1--10 keV band. We also retrieved the {\it Swift-BAT} 70 Month Hard X-ray Survey spectra of VII\,Zw\,73 and IRAS\,16288+3929 in the 15--100\,keV band.  
From the spectral analysis of the data we have obtained the following results:
\begin{enumerate}
\item NGC\,613, VII\,Zw\,73, and IRAS\,16288+3929, host highly absorbed but Compton-thin AGN, with measured column densities of 4--6$\times$10$^{23}$\,cm$^{-2}$;
\item UGC\,3789 and NGC\,6264, the two confirmed disk-maser galaxies, are Compton-thick ($N_{\rm H}>10^{24}$\,cm$^{-2}$), as inferred by the large $EW$ of their Fe K$\alpha$ lines and confirmed by diagnostic diagrams;
\item among all the extragalactic water maser sources associated with AGN, in which the absorbing column density could be estimated, 96\% (45/47) and 57\% (27/47) host highly-obscured and Compton-thick AGN, respectively;
\item when considering only disk maser galaxies, 86\% (18/21) of them host Compton-thick AGN
\end{enumerate}

Globally, this work clearly indicates and confirms the trend of water maser sources associated with AGN to show high column densities and, in particular, the strong connection between H$_2$O masers in accretion disks and Compton-thick AGN.

In addition, our analysis, benefiting from a database nearly twice as large as that of previous publications, confirms the correlation between the bolometric luminosity and accretion disk radius reported by \citet{tilak08} with a higher degree of confidence. Furthermore, we found that the anomalous position of NGC\,2273 in the $R_{\rm H2O}-L_{\rm Bol}$ plane might indicate the presence of clumpy material at the inner radius of the masing disk.


\section*{Acknowledgments}
F. P. acknowledges support by {\it INTEGRAL} ASI I/033/10/0 and ASI/INAF I/009/10/0. P. C. would like to thank Lincoln Greenhill and Avanti Tilak for helpful suggestions in the early stages of this project and Annika Kreikenbohm and Eugenia Litzinger for critically reading the manuscript. 

{99}

\label{lastpage} 

\end{document}